# Low Leakage Ferroelectric Heteroepitaxial Al$_{0.7}$Sc$_{0.3}$N Films on GaN

*Keisuke Yazawa[1,2]\*, Charles Evans[3], Elizabeth Dickey[3], Brooks Tellekamp[1], Geoff L. Brennecka[2] and Andriy Zakutayev[1]\**

1. Materials Science Center, National Renewable Energy Laboratory, Golden, Colorado 80401, United States

2. Department of Metallurgical and Materials Engineering, Colorado School of Mines, Golden, Colorado 80401, United States

3. Department of Materials Science and Engineering, Carnegie Mellon University, Pittsburgh, Pennsylvania 15213, United States

**Corresponding Authors**

\*E-mail: Keisuke.Yazawa@nrel.gov and Andriy.Zakutayev@nrel.gov

**Abstract**

Wurtzite (Al,Sc)N ferroelectrics are attractive for microelectronics applications due to their chemical and epitaxial structural compatibility with wurtzite semiconductors such as GaN and (Al,Ga)N. However, the leakage current in epitaxial stacks reported to date should be reduced for reliable device operation. Following the tradition of other semiconductor heterostructures, crystalline structural quality—as measured by breadth of diffraction peaks and correlating with dislocation density—is commonly used as a proxy for leakage current, but we demonstrate here that the crystalline mosaicity that dominates the broadening of diffraction peaks in epitaxial $Al_{0.7}Sc_{0.3}N$ stacks does not dominate leakage current. We report here well-saturated ferroelectric hysteresis loops and orders of magnitude lower leakage current (0.07 A cm$^{-2}$) compared to values reported in literature (1 ~ 19 A cm$^{-2}$) for sputter-deposited epitaxial $Al_{0.7}Sc_{0.3}N$/GaN of comparable crystalline quality to prior reports. Further, we show $Al_{0.7}Sc_{0.3}N$ on lattice-matched InGaN buffers with improved structural characteristics exhibits increased leakage characteristics. This demonstration and understanding can help to guide further efforts towards reliable wurtzite ferroelectric devices and prioritize approaches targeting further leakage current reduction.

**1. Introduction**

Switchable spontaneous polarization of ferroelectric materials has enabled unique device capabilities in microelectronics such as non-volatile ferroelectric memories (FeRAM, FeFET) and microelectromechanical (MEMS) systems [1–3]. The desired material properties and characteristics vary depending on application, but process compatibility to existing semiconductor fabrication lines, large and robust remanent polarization, and low leakage current are preferable to realize reliable and cost-competitive devices. Recent efforts have developed emerging ferroelectrics such as fluorites based on modified HfO$_2$ [4,5], wurtzite materials based on modified AlN [6–10] and ZnO [11,12], and perovskite nitrides. [13] These vibrant on-going studies on ferroelectric materials has potential to lead to next generation microelectronics.

Among ferroelectrics, wurtzite nitrides are promising candidates for such devices due to their enhanced piezoelectric and ferroelectric properties and process compatibility with both commercially relevant III-N semiconductors and Si. Wurtzite (Al,Sc)N shows an improved piezoelectric response compared to pristine AlN, [14] robust ferroelectricity with large spontaneous polarization, [6,8] and sub-microsecond polarization switching [15,16]. Indeed, various devices using wurtzite ferroelectrics such as ferroelectric tunnel junctions [17], ferroelectric HEMTs [18], and two dimensional FETs [19] have been demonstrated. Wurtzite nitrides are suitable for those devices due to the low deposition process temperature (< 400 °C) compatible with Si microelectronics, and lack of relatively mobile oxygen ion that causes fatigue and associated reliability concerns in many oxide ferroelectrics [20,21].

Leakage current of wurtzite nitride ferroelectrics is a critical issue for memory applications, allowing undesired charge injection through the ferroelectric gate [18]. Many studies on this material class show non-saturated polarization-electric hysteresis loops due to leakage current contributions, especially for epitaxial (Al,Sc)N/GaN stacks reported in literature [22–25] that are essential for a ferroelectric HEMT [18]. It is especially surprising that the leakage currents reported for such epitaxial stacks are far greater than those reported for non-epitaxial stacks of similar composition [6,16,22–26]. The origins of leakage current in (Al,Sc)N is not fully understood yet, but dislocation density, [27] impurity associated with Sc source [28], and nitrogen vacancy formation leading to the change in metal-insulator barrier height [29]. In addition, deep level defects [30,31], impurity segregation [31], and interface roughness [32] are known to contribute to leakage current. Thus, measurements of structural quality are commonly used as proxies for electrical quality.

In this study, we demonstrate low leakage current in 250 and 125 nm thick epitaxial $Al_{0.7}Sc_{0.3}N$ films on GaN-templated $Al_2O_3$ substrates. The leakage current of our $Al_{0.7}Sc_{0.3}N$ films on the GaN template is 0.07 and 0.15 A cm$^{-2}$ for 250 and 125 nm thick films at the coercive field, which is one to two orders of magnitude lower compared to previously-reported values (1-19 A cm$^{-2}$) [22,23,25]. Based on the corresponding structure and property comparison, crystallographic mosaicity is not the dominant parameter that controls the leakage current in (Al,Sc)N/GaN stacks reported in literature. $Al_{0.7}Sc_{0.3}N$ film on lattice-matched (In,Ga)N buffers, which shows better mosaicity yet larger leakage current, supports the crystal quality insensitivity to leakage current. We suggest that focusing on factors other than macroscopic crystal lattice quality such as impurities, microstructure, and/or interface roughness enables further improvements on the leakage currents to achieve reliable (Al,Sc)N/GaN microelectronic device operation.

2. Results

The crystallographic structure and texture of the $Al_{0.7}Sc_{0.3}N$ film deposited on GaN template was analyzed using x-ray diffraction (XRD). Fig. 1(a) shows θ-2θ scan profiles for the films and Fig. 1(b) shows ω rocking curves for the $Al_{0.7}Sc_{0.3}N$ (0002) diffraction peaks. Besides the diffraction peaks associated with the substrate and template, only the (0002) wurtzite peak is observed, indicating the film is highly textured out of plane. The full width half maximum (FWHM) angles of the rocking curves are 1.18° for $Al_{0.7}Sc_{0.3}N$ film, comparable to values reported for other films deposited by sputtering [23]. The rocking curve of off-axis (10$\bar{1}$2) reflection, which is sensitive to edge dislocations (supplementary Fig. S1), is comparable with the rocking curve of the surface normal (0002) diffraction, which is sensitive to screw dislocation. The epitaxial relationship is identified with XRD $\phi$ scan for off-axis ($10\bar{1}3$) diffraction peaks (Fig. 1(c)). The film diffraction peak is aligned with the peak from the underlying GaN, indicating a hexagonal-on-hexagonal epitaxial relationship (Fig. 1d). The GaN templates are rotated 30 degrees with respect to the sapphire substrates as expected from the wurtzite/sapphire epitaxial growth mechanism [33]. The wurtzite lattice parameters shown in Fig. 1d ($a$ = 3.22 Å and $c$ = 5.03 Å) are measured with reciprocal space maps (supplementary Fig. S2). The reciprocal space maps also show that the films are fully relaxed, accommodating the 1 % in-plane lattice mismatch between the $Al_{0.7}Sc_{0.3}N$ and GaN template.

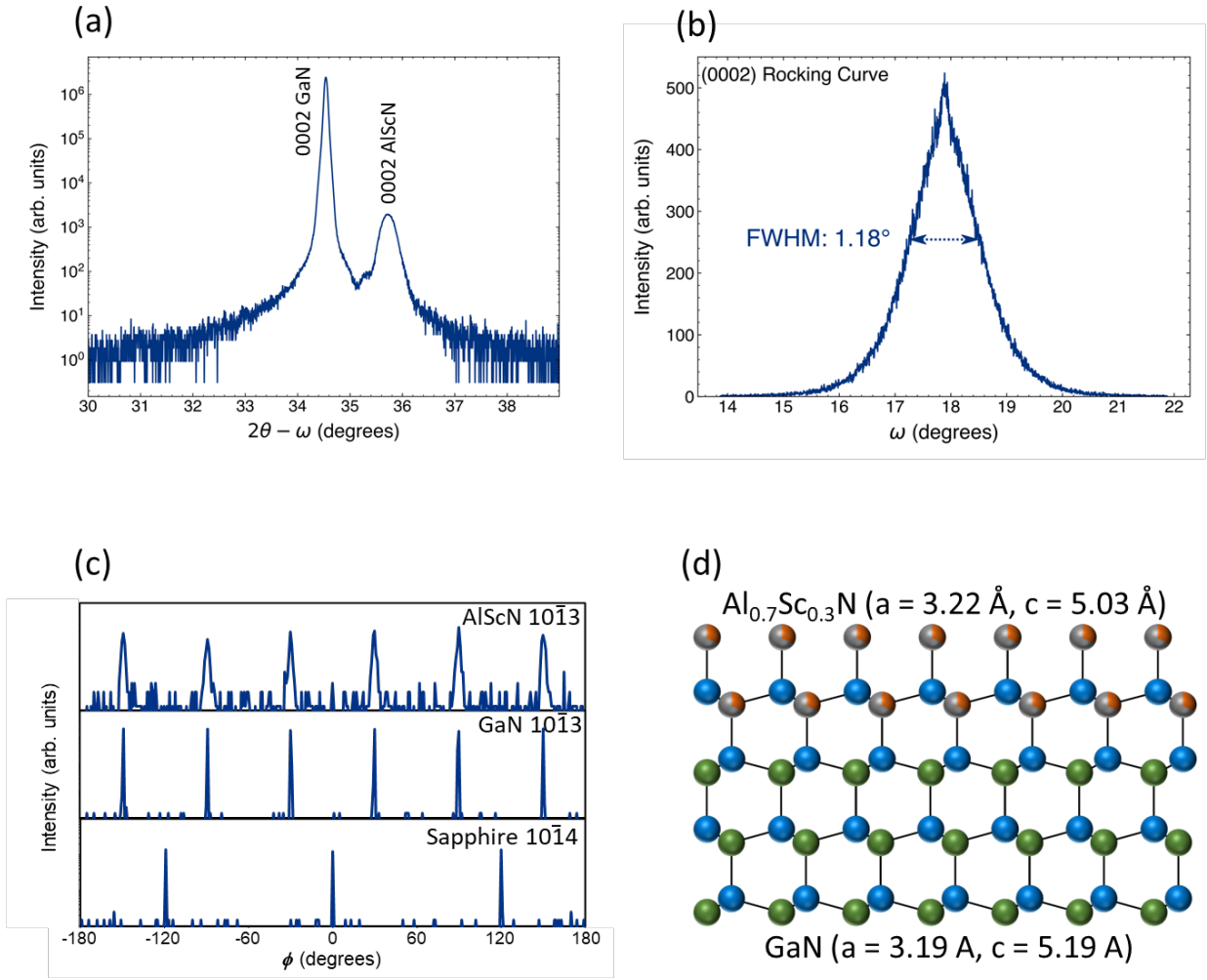

**Figure 1**. XRD crystallographic analysis of $Al_{0.7}Sc_{0.3}N$ films on GaN templated sapphire $Al_2O_3$ substrates. (a) 2θ–ω profiles and (b) omega rocking curves showing 0002 textured $Al_{0.7}Sc_{0.3}N$ wurtzite phase. (c) Off-axis diffraction XRD phi scan and (d) epitaxial relationship for $Al_{0.7}Sc_{0.3}N$ films on GaN film in the coherent approximation, showing film and template are in hexagonal-on-hexagonal relationship.

The microstructure of the samples was investigated using scanning transmission microscopy (STEM). Differential phase contrast (DPC) STEM (see methods for details), which is sensitive to small changes in sample orientation [34], was used to highlight the relative misorientation between neighboring crystallites. These images are shown in Figure 2 (a). The $Al_{0.7}Sc_{0.3}N$ film shows color contrast associated with crystal mosaicity, which is in good agreement with the rocking curve discussed above. The sharp chemical interface and wurtzite phase are confirmed with energy dispersed x-ray spectroscopy and selected area electron diffraction, respectively (supplementary Fig. S3). The similar densities of the screw and edge dislocations can qualitatively be seen in TEM images (supplementary Fig. S5), consistent with the XRD measurement results (Fig.1, Fig. S1)

Atomic resolution STEM imaging was also employed to examine the film-substrate interface; these images are shown in Fig. 2 (b) and Fig. 2 (c). The annular dark field (ADF) images – Fig. 2 (b) – show strong atomic number (Z) contrast between the relatively lighter elements present in the film and the heavier elements present in the substrate, and as such highlight the location of the interface. Along with the ADF images, simultaneously acquired dDPC images – Fig. 2 (c) – can be used to visualize both light and heavy elements. These images confirm that in both cases the films were grown in a nitrogen-polar orientation.

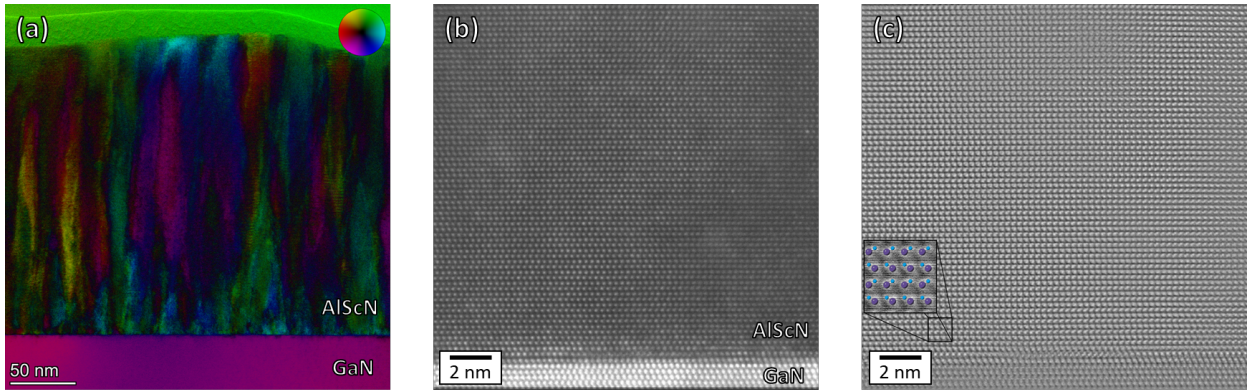

**Figure 2**. STEM images and analysis of the $Al_{0.7}Sc_{0.3}N$ films grown on GaN. (a) Differential phase contrast images highlighting the relative misorientation between crystallites for the film. (b) Atomic resolution ADF images highlighting the film/template interface. (c) dDPC images acquired simultaneously with the respective ADF images.

The nested polarization-electric field loops, taken with 1 kHz triangular applied electric field, show well-saturated hysteresis indicated by their square shape (Fig. 3a). The leakage current is directly measured with the current – electric field loop (Fig. 3b). To observe leakage current contributions clearly, the capacitive current density, $j_c$ is subtracted in the plot (See Methods). Besides the leakage current contribution, the sharp current peaks associated with ferroelectric switching at the coercive field are observed.

The leakage current at the coercive field directly correlates to energy loss and charge injection during switching for ferroelectric device applications. Based on Fig. 3b, the leakage current densities at the positive and negative respective coercive fields are 0.13 A cm$^{-2}$ and 0.02 A cm$^{-2}$ for the film on GaN template, and are much smaller compared to the polarization current (>10-20 A cm$^{-1}$) The leakage current of 0.01 A cm$^{-1}$ under an electric field < 4000 kV cm$^{-1}$ is the noise floor of the measurement equipment.

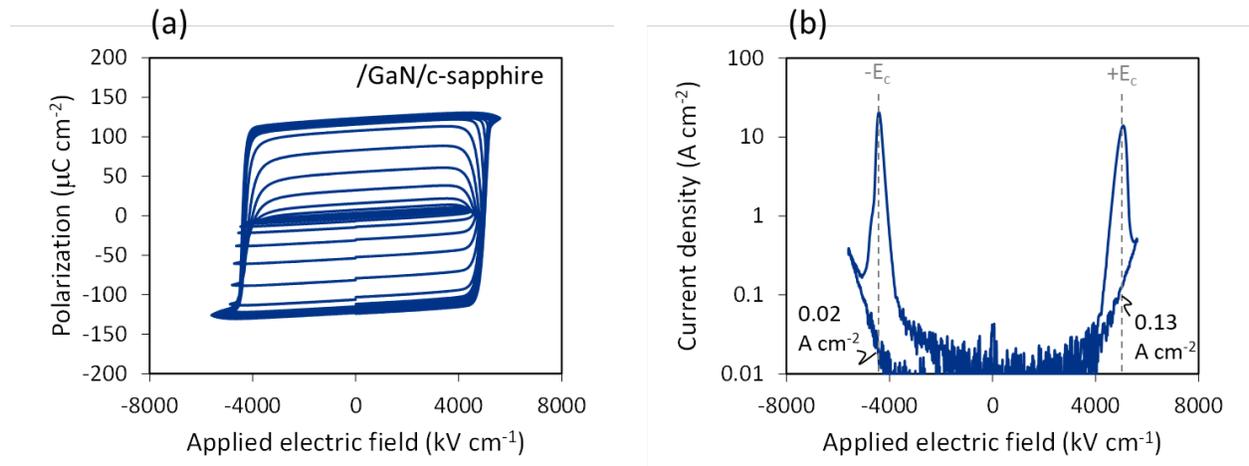

**Figure 3**. Ferroelectric properties characterization for Al$_{0.7}$Sc$_{0.3}$N films on GaN templated substrates. (a) Nested ferroelectric hysteresis loops showing the polarization saturation. Excitation triangular field frequencies are 1 kHz for film on GaN. (b) Current loops showing switching current and leakage current.

A thinner film (125 nm) deposited using identical growth conditions but half the deposition time is also investigated since thickness also affects leakage current. [35,36] The ferroelectric hysteresis loop is well-saturated, and the leakage current densities at the positive and negative respective coercive fields are 0.20 A cm$^{-2}$ and 0.11 A cm$^{-2}$ (Supplementary Fig. S4). Those slightly larger leakage current densities compared to the 250 nm film are in good agreement with the leakage increase trend with thickness reduction [35], but still orders of magnitude lower than the reported leakage current of (Al,Sc)N with various thickness (100-300 nm). [22,23,25]

## 3. Discussion

Both the 250 and 125 nm films on GaN template studied here showed significantly lower leakage current than that of previously-reported (Al,Sc)N/GaN films. Table 1 compares properties of ferroelectric (Al,Sc)N/GaN films reported to date, including this study. The films on GaN template in this study show one to two orders of magnitude lower leakage current than the other reported films. The leakage current density listed in Table 1 is an average value of the current under +$E_c$ and -$E_c$. The asymmetry is attributed to two convoluted factors: asymmetric leakage current due to the different electrode contacts (n-GaN or Pt) and asymmetric $E_c$ due to trapped charges [37–39] and/or polarity dependence on wake-up behavior [40]. This low leakage current $Al_{0.7}Sc_{0.3}N$/GaN film obtained in this study is promising for microelectronics device applications.

Rocking curve FWHM is a reliable proxy measurement for structural quality because it is sensitive to dislocation densities [41], and it provides a convenient comparison of the structural quality across films from different studies. The rocking curve FWHM of the $Al_{0.7}Sc_{0.3}N$ films is comparable to a reported sputter deposited (Al,Sc)N/GaN film [23] and wider than MBE or MOCVD films. This comparison provides an insight into the fact that epitaxial quality is not the dominant parameter for the reported leakage current. There are many possible key differences affecting the leakage current such as concentration of nitrogen vacancies, oxygen content, interface roughness, etc. Those data are not revealed across the studies, so that it is challenging to carry out direct comparison and understand the root cause of leakage current differences.

**Table 1**. Property comparison of epitaxial (Al,Sc)N film on GaN template

|  | This study |  | Schönweger [23] |  | Wang [22] | Wolff [25] |
| --- | --- | --- | --- | --- | --- | --- |
| Sc content x | 0.3 |  | 0.28 |  | 0.21 | 0.15 |
| Thickness | 250 nm | 125 nm | 300 nm | 100 nm | 100 nm | 230 nm |
| Template | n-GaN | n-GaN | n-GaN | n-GaN | n-GaN | n-GaN |
| $E_c$ | 4.7 MV cm$^{-1}$ @ 1 kHz | 5.1 MV cm$^{-1}$ @ 1 kHz | 4.6 MV cm$^{-1}$ @ 1.5 kHz | 5.5 MV cm$^{-1}$ @ 2 kHz | 4.6 MV cm$^{-1}$ @ 10 kHz | 5.5 MV cm$^{-1}$ @ 1.5 kHz |
| Leakage current at $E_c$ | **0.07 A cm$^{-2}$** | **0.15 A cm$^{-2}$** | 1 A cm$^{-2}$ | 4 A cm$^{-2}$ | 19 A cm$^{-2}$ | 8 A cm$^{-2}$ |
| Device yield | 8/10 | 7/10 | - | - | - | - |
| (0002) rocking curve FWHM | 1.18° | 1.27° | ~1° | ~1° | ~0.4° | 0.07° |
| Device size | 50 mm in diameter |  | 100 mm × 100 mm | 50 mm × 50 mm | 20 mm in diameter | 20 mm in diameter |
| Deposition technique | RF sputtering |  | Pulsed DC sputtering |  | MBE | MOCVD |

For further understanding on the rocking curve – leakage current relationship, we grew films on In$_{0.18}$Ga$_{0.82}$N templates with a better lattice match to Al$_{0.7}$Sc$_{0.3}$N (Fig. 4a). As expected from this more favorable epitaxial relation, these Al$_{0.7}$Sc$_{0.3}$N/In$_{0.18}$Ga$_{0.82}$N stacks exhibit a narrower rocking curve FWHM (= 0.94°) and larger grain size (Fig. 4b, 4c and supplementary Fig. S6). However, the Al$_{0.7}$Sc$_{0.3}$N/In$_{0.18}$Ga$_{0.82}$N film possesses a significantly rougher interface (1.92 ± 0.40 nm) than the Al$_{0.7}$Sc$_{0.3}$N/GaN film (0.41 ± 0.19 nm) as seen in Fig. 4d (compare with figure 2b), due to limitations in the In$_{0.18}$Ga$_{0.82}$N sample growth. Thus, in addition to providing the better lattice match, the In$_{0.18}$Ga$_{0.82}$N buffer introduce another uncontrolled variable (roughness).

Leakage current, observed as a polarization rounding (lack of saturation) in the hysteresis loops and current loop of the film on In$_{0.18}$Ga$_{0.82}$N as shown in Fig. 4e and Fig. 4f respectively. The leakage current densities at the positive and negative respective coercive fields are 4.7 A cm$^{-2}$ and 0.6 A cm$^{-2}$, which is an order of magnitude larger than the Al$_{0.7}$Sc$_{0.3}$N film on GaN shown in Fig. 3b and Table 1. This leakage current increase compared to Al$_{0.7}$Sc$_{0.3}$N/GaN in this study is likely attributed to the interface roughness that is known to be a key factor to increase leakage current [32]. In other words, the leakage reduction from the narrower rocking curve is less significant than the roughness contribution. Hence, understanding the leakage reduction mechanisms will require further focused study.

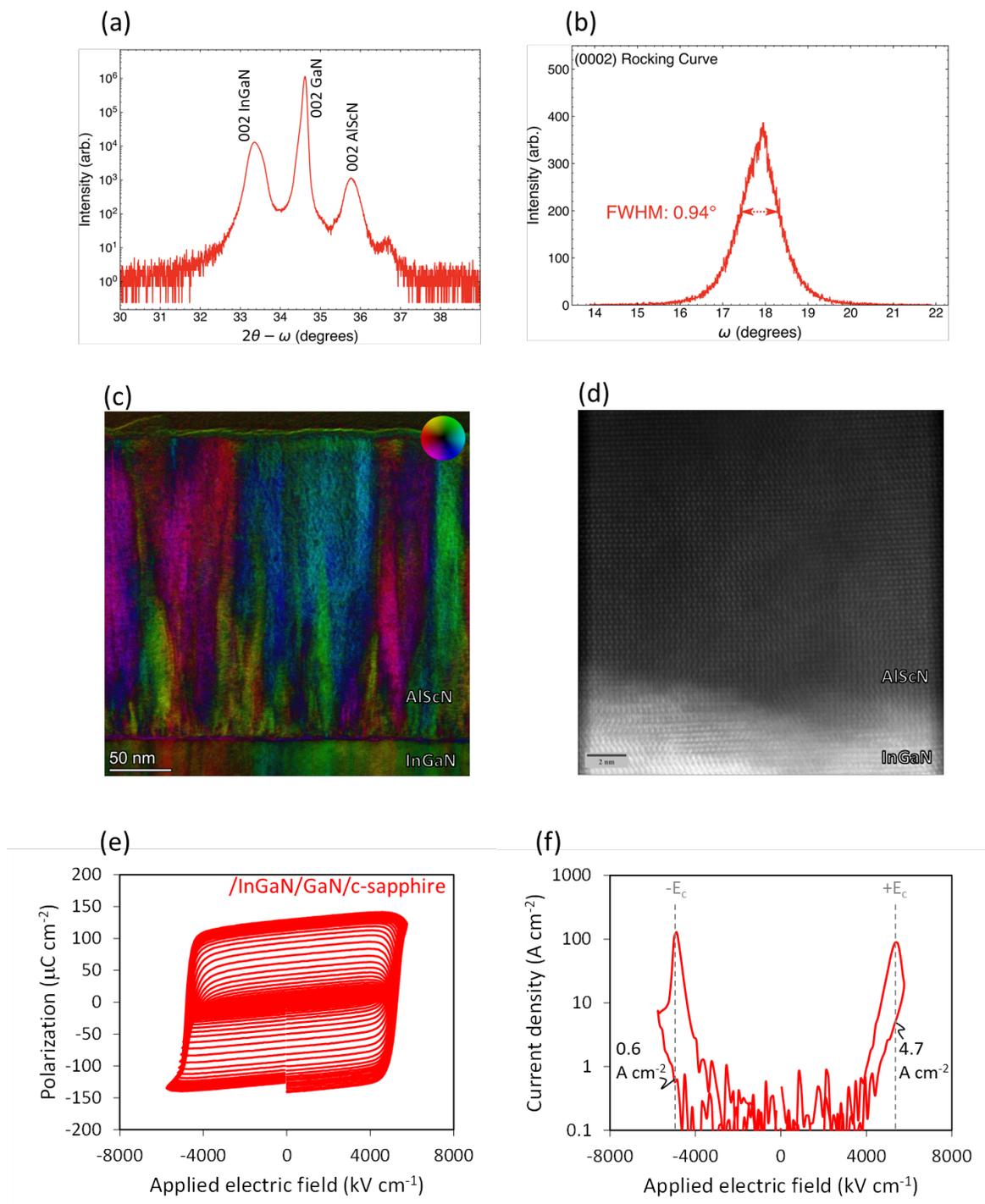

**Figure 4**. Characterization of $Al_{0.7}Sc_{0.3}N$ films on lattice matched $In_{0.18}Ga_{0.82}N$ templated substrates. (a) $2\theta-\omega$ profiles and (b) omega rocking curve scans showing 0002 textured $Al_{0.7}Sc_{0.3}N$ wurtzite phase. (c) Differential phase contrast image highlighting the relative misorientation between crystallites for the film. (d) Atomic resolution ADF images highlighting the film/template interface. (e) Nested ferroelectric hysteresis loops with excitation triangular field frequencies are 10 kHz. (f) Current loops showing switching current and leakage current.

## 4. Conclusion

We demonstrate epitaxial $Al_{0.7}Sc_{0.3}N$/GaN thin film deposited via RF reactive magnetron sputtering on sapphire substrates exhibits low leakage currents, 0.07 A cm$^{-2}$, which is a few orders of magnitude lower than reported for epitaxial (Al,Sc)N/GaN stacks in literature. The hexagon-on-hexagon epitaxial relationship is confirmed by X-ray and electron diffraction. The X-ray rocking curve FWHM is 1.18°, which is comparable to those reported for sputtered (Al,Sc)N films with ~10x larger leakage currents, and 3 times wider than MBE or MOCVD deposited film which show ~100x larger leakage currents. This indicates the crystal quality of (Al,Sc)N film is not the dominant factor of the leakage current reported in (Al,Sc)N/GaN stacks. Larger leakage current of narrower FWHM (0.94°) $Al_{0.7}Sc_{0.3}N$ grown on lattice-matched (In,Ga)N buffer layers is likely attributed to increased interface roughness, which supports the relative insensitivity of crystal quality to leakage current. These results demonstrate progress in mitigating the leakage current of (Al,Sc)N/GaN and provide insights for material design with improved energy efficiency and memory retention in ferroelectric microelectronics devices.

**Acknowledgements**

This work was co-authored by Colorado School of Mines and the National Renewable Energy Laboratory, operated by the Alliance for Sustainable Energy, LLC, for the U.S. Department of Energy (DOE) under Contract No. DE-AC36-08GO28308. Funding was provided by the by the Office of Science (SC), Office of Basic Energy Sciences (BES) as part of the Early Career Award "Kinetic Synthesis of Metastable Nitrides" (AlScN material synthesis), by the National Science Foundation under Grant No. DMR-2119281 (TEM observation, electrical characterization, and data interpretation), and by the SC, BES under program ERW6548 (crystal structural analysis). The authors acknowledge use of the Materials

Characterization Facility at Carnegie Mellon University supported by grant MCF-677785. Some of the work was performed in part of Colorado School of Mines' Shared Instrumentation Facility (RRID:SCR_022053). The data affiliated with this study are available from the corresponding author upon reasonable request. The views expressed in the article do not necessarily represent the views of the DOE or the U.S. Government.

**Conflict of Interest**

The authors declare no conflict of interest.

**Reference**

[1]  J. F. Scott, C. A. Paz, and D. Araujo, Source Sci. New Ser. **246**, 1400 (1989).

[2]  J. L. Moll and Y. Tarui, IEEE Trans. Electron Devices **10**, 338 (1963).

[3]  S. Trolier-Mckinstry and P. Muralt, J. Electroceramics **12**, 7 (2004).

[4]  T. S. Böscke, J. Müller, D. Bräuhaus, U. Schröder, and U. Böttger, Appl. Phys. Lett. **99**, (2011).

[5]  J. Müller, T. S. Böscke, D. Bräuhaus, U. Schröder, U. Böttger, J. Sundqvist, P. Kcher, T. Mikolajick, and L. Frey, Appl. Phys. Lett. **99**, 112901 (2011).

[6]  S. Fichtner, N. Wolff, F. Lofink, L. Kienle, and B. Wagner, J. Appl. Phys. **125**, 114103 (2019).

[7]  S. Yasuoka, T. Shimizu, A. Tateyama, M. Uehara, H. Yamada, M. Akiyama, Y. Hiranaga, Y. Cho, and H. Funakubo, J. Appl. Phys. **128**, 114103 (2020).

[8]  K. Yazawa, D. Drury, A. Zakutayev, and G. L. Brennecka, Appl. Phys. Lett. **118**, 162903 (2021).

[9]  J. Hayden, M. D. Hossain, Y. Xiong, K. Ferri, W. Zhu, M. V. Imperatore, N. Giebink, S. Trolier-Mckinstry, I. Dabo, and J. P. Maria, Phys. Rev. Mater. **5**, 044412 (2021).

[10] D. Wang, S. Mondal, J. Liu, M. Hu, P. Wang, S. Yang, D. Wang, Y. Xiao, Y. Wu, T. Ma, and Z. Mi, Appl. Phys. Lett. **123**, (2023).

[11] K. Ferri, S. Bachu, W. Zhu, M. Imperatore, J. Hayden, N. Alem, N. Giebink, S. Trolier-McKinstry, and J.-P. Maria, J. Appl. Phys. **130**, 044101 (2021).

[12] R. Ogawa, A. Tamai, K. Tanaka, H. Adachi, and I. Kanno, Jpn. J. Appl. Phys. **63**, 010902 (2023).

[13] K. R. Talley, C. L. Perkins, D. R. Diercks, G. L. Brennecka, and A. Zakutayev, Science (80-. ). **374**, 1488 (2021).

[14] M. Akiyama, T. Kamohara, K. Kano, A. Teshigahara, Y. Takeuchi, and N. Kawahara, Adv. Mater. **21**, 593 (2009).

[15] D. Wang, P. Musavigharavi, J. Zheng, G. Esteves, X. Liu, M. M. A. Fiagbenu, E. A. Stach, D. Jariwala, and R. H. Olsson, Phys. Status Solidi - Rapid Res. Lett. **15**, 2000575 (2021).

[16] K. Yazawa, J. Hayden, J. Maria, W. Zhu, S. Trolier-McKinstry, A. Zakutayev, and G. L. Brennecka, Mater. Horizons (2023).

[17] X. Liu, J. Zheng, D. Wang, P. Musavigharavi, E. A. Stach, R. Olsson III, and D. Jariwala, (2020).

[18] D. Wang, P. Wang, M. He, J. Liu, S. Mondal, M. Hu, D. Wang, Y. Wu, T. Ma, and Z. Mi, Appl. Phys. Lett. **122**, 090601 (2023).

[19] K.-H. Kim, S. Oh, M. Mercy, A. Fiagbenu, J. Zheng, P. Musavigharavi, P. Kumar, N. Trainor, A. Aljarb, Y. Wan, M. Kim, K. Katti, S. Song, G. Kim, Z. Tang, J.-H. Fu, M. Hakami, V. Tung, J. M. Redwing, E. A. Stach, R. H. Olsson Iii, and D. Jariwala, Nat. Nanotechnol. 2023 1 (2023).

[20] S. Mueller, J. Müller, U. Schroeder, and T. Mikolajick, IEEE Trans. Device Mater. Reliab. **13**, 93 (2013).

[21] S. J. Yoon, D. H. Min, S. E. Moon, K. S. Park, J. Il Won, and S. M. Yoon, IEEE Trans. Electron Devices **67**, 499 (2020).

[22] P. Wang, D. Wang, N. M. Vu, T. Chiang, J. T. Heron, and Z. Mi, Appl. Phys. Lett. **118**, 223504 (2021).

[23] G. Schönweger, A. Petraru, R. Islam, N. Wolff, B. Haas, A. Hammud, C. Koch, L. Kienle, H. Kohlstedt, S. Fichtner, G. Schönweger, A. Petraru, H. Kohlstedt, M. R. Islam, N. Wolff, L. Kienle, S. Fichtner, B. Haas, C. Koch, and A. Hammud, Adv. Funct. Mater. **32**, 2109632 (2022).

[24] P. Wang, D. Wang, S. Mondal, and Z. Mi, Appl. Phys. Lett. **121**, 023501 (2022).

[25] N. Wolff, G. Schönweger, I. Streicher, M. R. Islam, N. Braun, P. Straňák, L. Kirste, M. Prescher, A. Lotnyk, H. Kohlstedt, S. Leone, L. Kienle, and S. Fichtner, Adv. Phys. Res. 2300113 (2024).

[26] K. Yazawa, J. S. Mangum, P. Gorai, G. L. Brennecka, and A. Zakutayev, J. Mater. Chem. C **10**, 17557 (2022).

[27] D. Wang, P. Wang, S. Mondal, Y. Xiao, M. Hu, and Z. Mi, Appl. Phys. Lett. **121**, 42108 (2022).

[28] J. Casamento, H. Lee, C. S. Chang, M. F. Besser, T. Maeda, D. A. Muller, H. (Grace) Xing, and D. Jena, APL Mater. **9**, (2021).

[29] J. Kataoka, S. L. Tsai, T. Hoshii, H. Wakabayashi, K. Tsutsui, and K. Kakushima, Jpn. J. Appl. Phys. **60**, 030907 (2021).

[30] N. U. Din, C.-W. Lee, G. L. Brennecka, and P. Gorai, (2023).

[31] A. Y. Polyakov and I. H. Lee, Mater. Sci. Eng. R Reports **94**, 1 (2015).

[32] Y. P. Zhao, G. C. Wang, T. M. Lu, G. Palasantzas, and J. T. M. De Hosson, Phys. Rev. B **60**, 9157 (1999).

[33] P. Kung, C. J. Sun, A. Saxler, H. Ohsato, and M. Razeghi, J. Appl. Phys. **75**, 4515 (1994).

[34] D. J. Taplin, N. Shibata, M. Weyland, and S. D. Findlay, Ultramicroscopy **169**, 69 (2016).

[35] S. L. Tsai, T. Hoshii, H. Wakabayashi, K. Tsutsui, T. K. Chung, E. Y. Chang, and K. Kakushima, Jpn. J. Appl. Phys. **60**, SBBA05 (2021).

[36] G. Schönweger, M. R. Islam, N. Wolff, A. Petraru, L. Kienle, H. Kohlstedt, and S. Fichtner, Phys. Status Solidi – Rapid Res. Lett. **17**, 2200312 (2023).

[37] A. K. Tagantsev and G. Gerra, J. Appl. Phys. **100**, (2006).

[38] M. Grossmann, O. Lohse, D. Bolten, U. Boettger, T. Schneller, and R. Waser, J. Appl. Phys. **92**, 2680 (2002).

[39] T. Friessnegg, S. Aggarwal, R. Ramesh, B. Nielsen, E. H. Poindexter, and D. J. Keeble, Appl. Phys. Lett. **77**, 127 (2000).

[40] K. Yazawa, D. Drury, J. Hayden, J. P. Maria, S. Trolier-McKinstry, A. Zakutayev, and G. L. Brennecka, J. Am. Ceram. Soc. (2023).

[41] J. E. Ayers, J. Cryst. Growth **135**, 71 (1994).

# Supplemental Material

# Low Leakage Ferroelectric Heteroepitaxial Al$_{0.7}$Sc$_{0.3}$N Films on GaN

*Keisuke Yazawa[1,2]\*, Charles Evans[3], Elizabeth Dickey[3], Brooks Tellekamp[1], Geoff L. Brennecka[2] and Andriy Zakutayev[1]\**

1. Materials Science Center, National Renewable Energy Laboratory, Golden, Colorado 80401, United States

2. Department of Metallurgical and Materials Engineering, Colorado School of Mines, Golden, Colorado 80401, United States

3. Department of Materials Science and Engineering, Carnegie Mellon University, Pittsburgh, Pennsylvania 15213, United States

**Corresponding Authors**

*E-mail: Keisuke.Yazawa@nrel.gov and Andriy.Zakutayev@nrel.gov

**Experimental Methods**

**Al$_{0.7}$Sc$_{0.3}$N deposition**

The 250 nm Al$_{0.7}$Sc$_{0.3}$N films were deposited on the (0001) oriented GaN-on-sapphire template and MBE deposited (In,Ga)N via reactive RF magnetron sputtering using the following growth conditions: 2 mTorr of Ar/N$_2$ (15/5 sccm flow), and a target power density of 6.6 W/cm$^2$ on a 2" diameter Al$_{0.7}$Sc$_{0.3}$ alloy target (Stanford Advanced Materials). The substrate was rotated and heated to 400 °C during deposition. The base pressure, partial oxygen and water vapor pressure at 400 °C were $< 2 \times 10^{-7}$ torr, $P_{O2} < 2 \times 10^{-8}$ torr and $P_{H2O} < 1 \times 10^{-7}$ torr, respectively. Top Au (100nm)/Ti (20 nm) contacts 50 µm in diameter were deposited on the Al$_{0.7}$Sc$_{0.3}$N film film via electron beam evaporation through a photolithographic pattern.

**In$_{0.18}$Ga$_{0.82}$N buffer deposition**

The InGaN buffer layer was grown in a Riber compact 21 Molecular Beam Epitaxy (MBE) system using previously published methods [1]. An (0001) oriented GaN-on-sapphire template (6 µm thick, Si-doped, 3LCorp) was used as a substrate. The GaN surface was first regrown using metal modulated epitaxy (MME), monitoring the III/V ratio by reflection high energy electron diffraction (RHEED) intensity transients [2,3]. The GaN regrowth was ~130 nm in thickness, after which the substrate thermocouple setpoint was lowered to 600 °C, and the III-V ratio was lowered to a slightly nitrogen rich regime (III/V = 0.92) targeting 18% In. The InGaN was grown to a total thickness of 100 nm. During growth the relaxed in-plane lattice constant was monitored by RHEED, and the In/Ga ratio was adjusted after relaxation to obtain an in-plane lattice constant of approximately 3.23 Å.

**XRD measurements**

Crystal structure was analyzed using a Rigaku Smartlab with Cu Kα radiation monochromated by a 2-bounce Ge (220) channel cut crystal and columnated with a parabolic mirror. (0002) rocking curves, used to analyze mosaic tilt, were obtained using a 0.5 mm slit on the diffracted beam path. (10$\bar{1}$2) rocking curves, used to analyze combined mosaic tilt and twist, were obtained in a skew-symmetric geometry by tilting χ using a 1 mm slit on the diverging beam path. [4]

**TEM imaging**

Scanning transmission electron microscopy (STEM) was conducted on a Thermo Fisher Themis 200 G3 aberration corrected microscope operating at 200 keV, with a probe semi-angle of convergence of 17.9 mrad. Differential phase contrast [5] (DPC) imaging was conducted using a four-segment annular detector, with a camera length corresponding to inner and outer collection angles of 11 and 43 mrad, respectively. Annular dark field (ADF) images (46–200 mrad inner and outer collection angles) were collected simultaneously with DPC. The roughness of the film-substrate interfaces was measured by marking the position of the interface across the width of each DPC image and then calculating the root-mean-squared deviation of the interface height from its average within each image. Roughness values measured in this way may not be comparable to roughness values measured with techniques owing to the projection of one of the interface dimensions in the STEM images, but are comparable with each other.

**Ferroelectric measurements**

Ferroelectric polarization – electric field hysteresis measurements, and current – electric field loop measurements, were taken with a Precision Multiferroic system from Radiant

Technologies. The applied triangle excitation field was up to 5.6 MV cm$^{-1}$ at 1 kHz for the Al$_{0.7}$Sc$_{0.3}$N on the GaN template, and 5.8 MV cm$^{-1}$ at 10 kHz for the Al$_{0.7}$Sc$_{0.3}$N on the In$_{0.18}$Ga$_{0.82}$N template.

To extract the leakage current contribution from the current – electric field loop, the capacitive current was subtracted. The capacitive current is defined as

$$j_c = \frac{1}{A}\frac{dQ}{dt} = \frac{CdV}{Adt} = \frac{\varepsilon_r \varepsilon_0}{d}\frac{dV}{dt}$$

where $A$ is the device area, $C$ is the capacitance, $\varepsilon_r$ is relative permittivity, $\varepsilon_0$ is the permittivity of vacuum, $d$ is the film thickness. This constant current in the triangular excitation (d$V$/d$t$ is constant) was subtracted from the raw current loop data.

**Supplementary Figures**

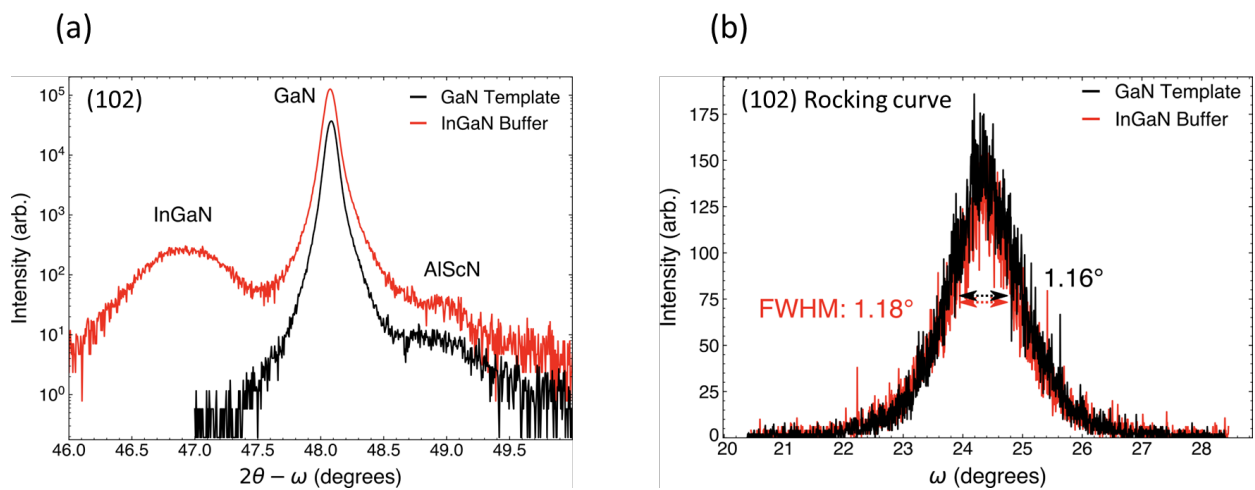

**Figure S1**. Off axis (102) XRD diffraction analysis. (a) 2θ-ω scan of $Al_{0.7}Sc_{0.3}N/GaN$ and $Al_{0.7}Sc_{0.3}N/In_{0.18}Ga_{0.82}N$. (b) (102) rocking curve of $Al_{0.7}Sc_{0.3}N$ for both samples. There is no significant difference in the values between the two samples, and the FWHM values are comparable to the (0002) rocking curves (1.18° and 0.94°).

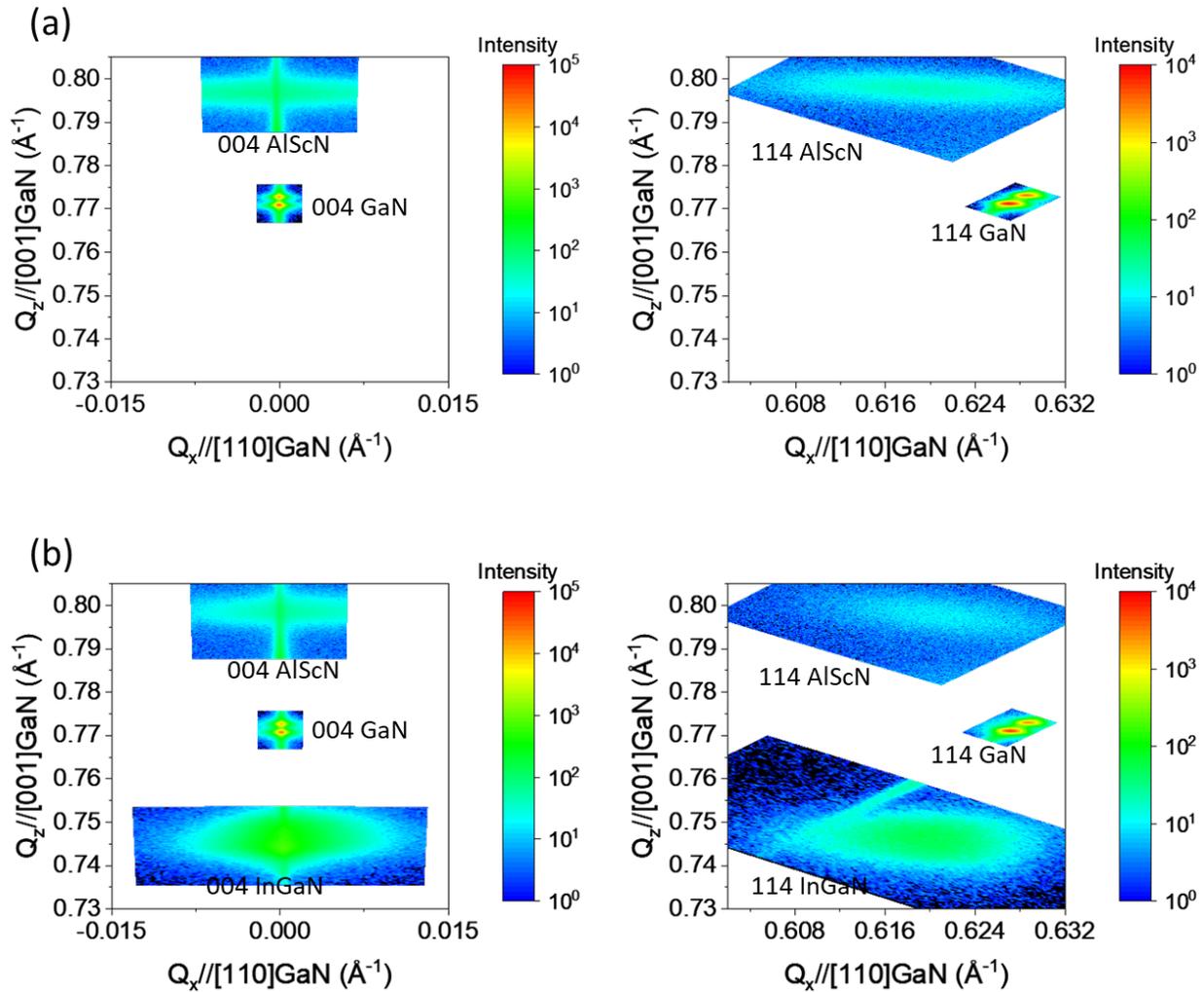

**Figure S2**. Reciprocal space maps for (a) $Al_{0.7}Sc_{0.3}N/GaN$ and (b) $Al_{0.7}Sc_{0.3}N/In_{0.18}Ga_{0.82}N$ films. From the peak positions, the *a* and *c* lattice parameters are determined to be 3.22 Å and 5.03 Å for $Al_{0.7}Sc_{0.3}N$ on GaN and 3.23 Å and 5.01 Å for $Al_{0.7}Sc_{0.3}N$ on $In_{0.18}Ga_{0.82}N$. The in-plane lattice parameter is matched for the $Al_{0.7}Sc_{0.3}N/In_{0.18}Ga_{0.82}N$ interface while a slight in-plane compressive strain is seen in $Al_{0.7}Sc_{0.3}N$ on GaN attributed to the lattice mismatch (a lattice parameter of GaN is 3.19 Å). The peaks include Cu k$\alpha_2$ diffraction.

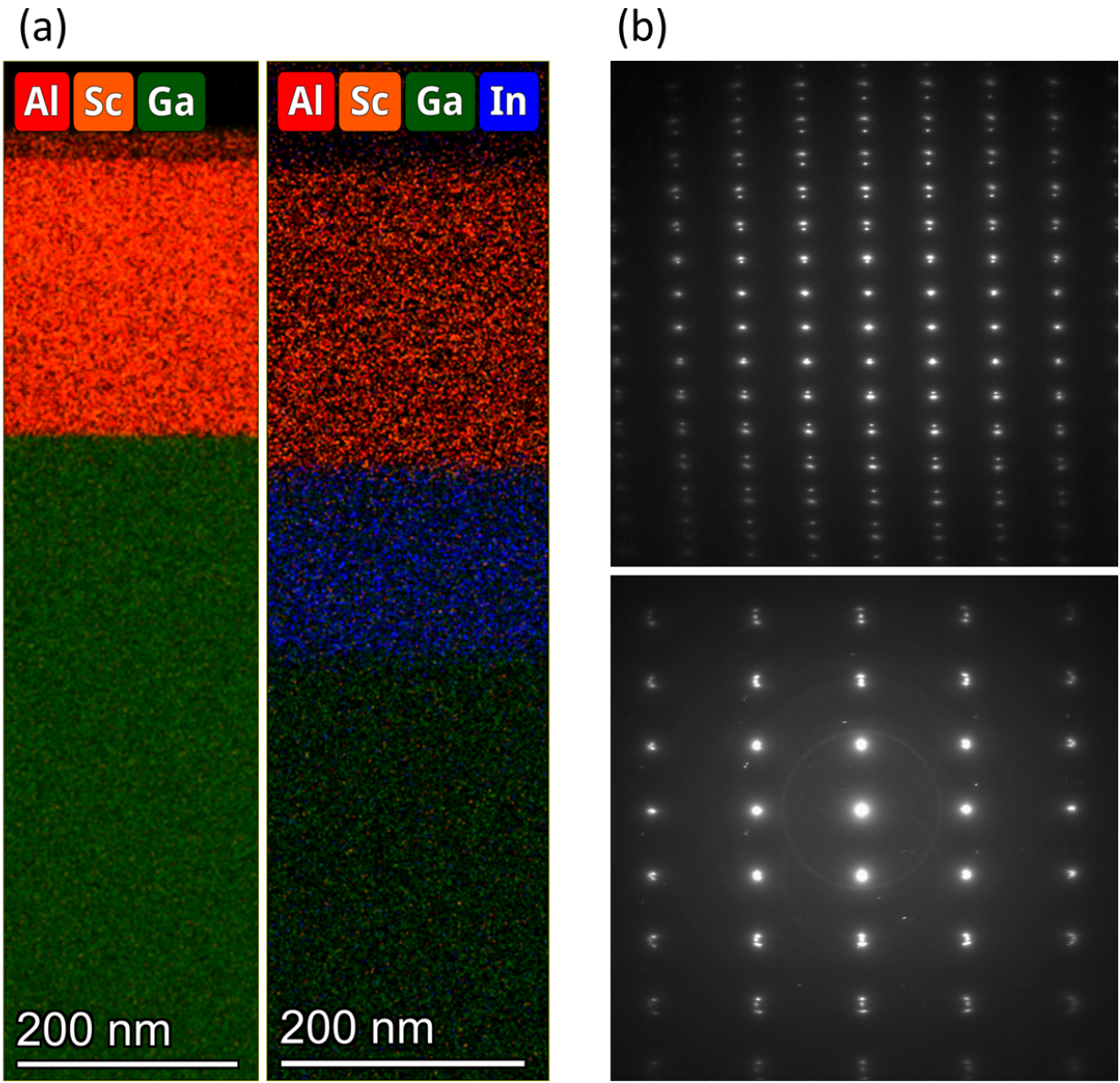

**Figure S3**. TEM analysis on $Al_{0.7}Sc_{0.3}N$ films. (a) Al, Sc, Ga and In elemental mappings via dispersed x-ray spectroscopy show sharp chemical interface and uniform chemical distribution. (b) Selected area electron diffraction shows the wurtzite symmetry.

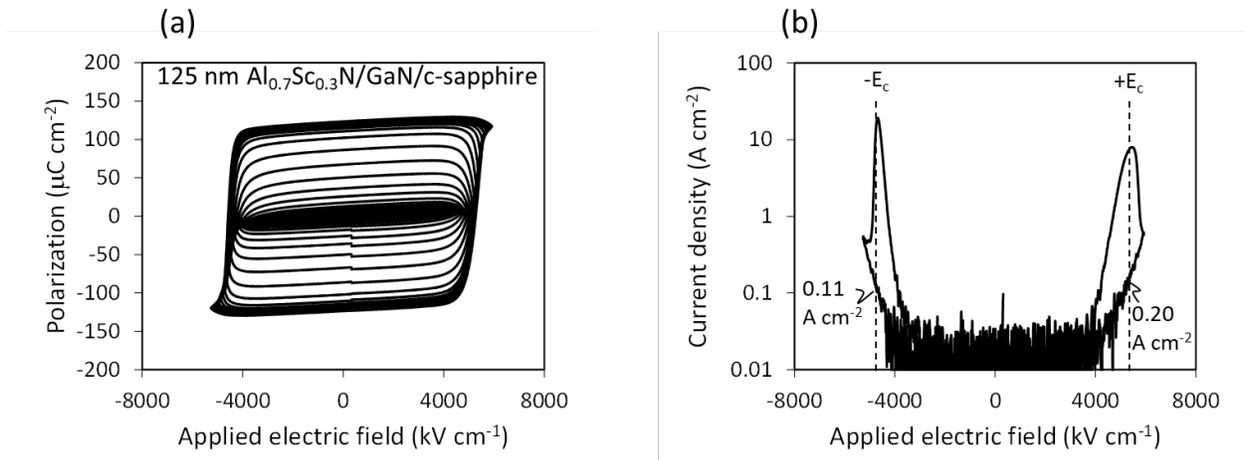

**Figure S4**. Ferroelectric properties characterization for 125 nm $Al_{0.7}Sc_{0.3}N$ films on GaN templated substrates. (a) Nested ferroelectric hysteresis loops showing the polarization saturation. Excitation triangular field frequencies are 1 kHz for film on GaN. (b) Current loops showing switching current and leakage current.

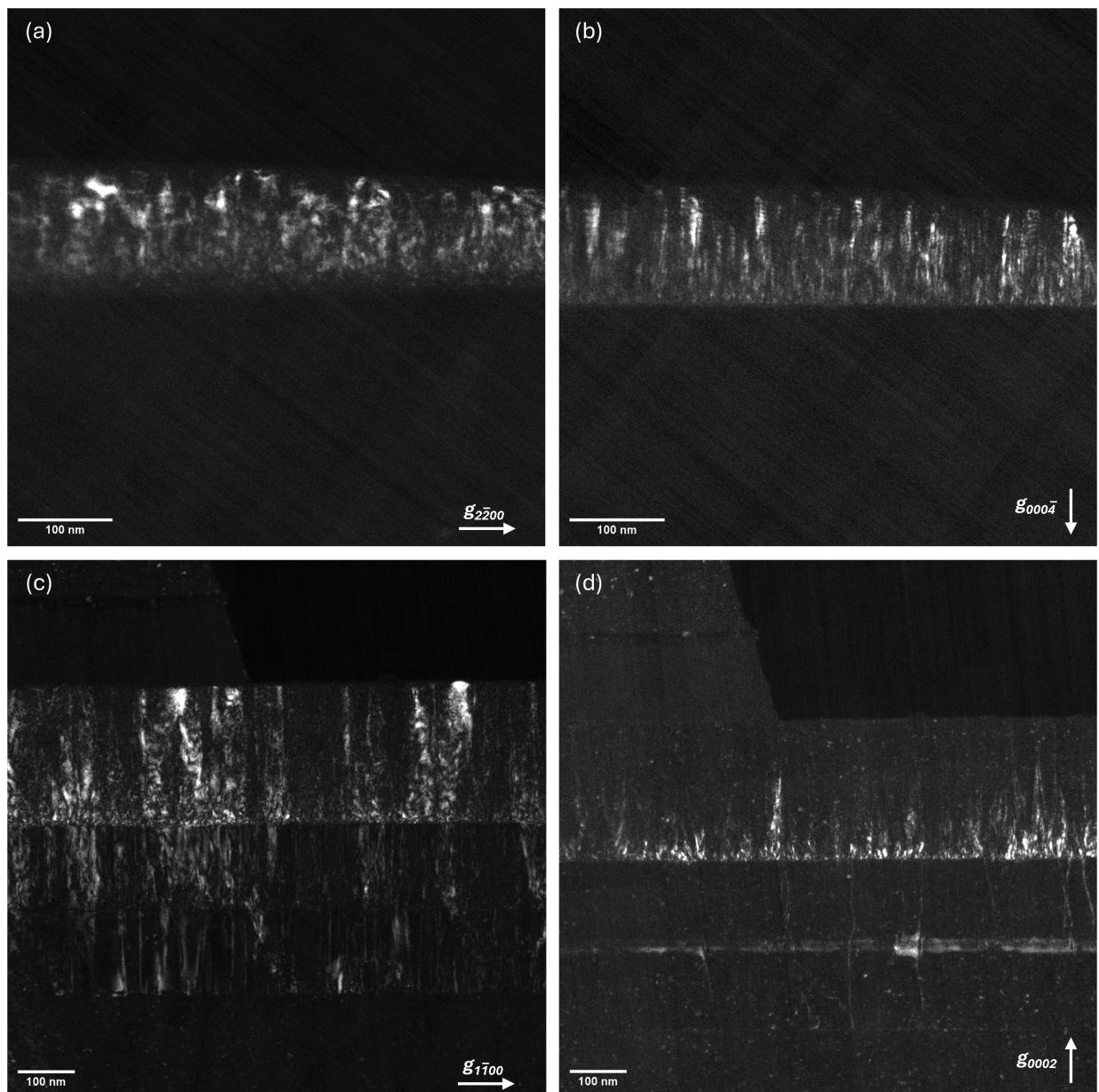

**Figure S5.** Weak-beam dark field images from Al0.7Sc0.3N films. (a) Grown on GaN, imaged using g=2$\bar{2}$00 and (b) g=000$\bar{4}$. (c) Grown on In0.18Ga0.82N, imaged using g=1$\bar{1}$00 and (d) g=0002. Contrast from edge dislocations can appear in (a) and (c), while contrast from screw dislocations can appear in (b) and (d). Some contrast due to mosaicity can be seen in all images.

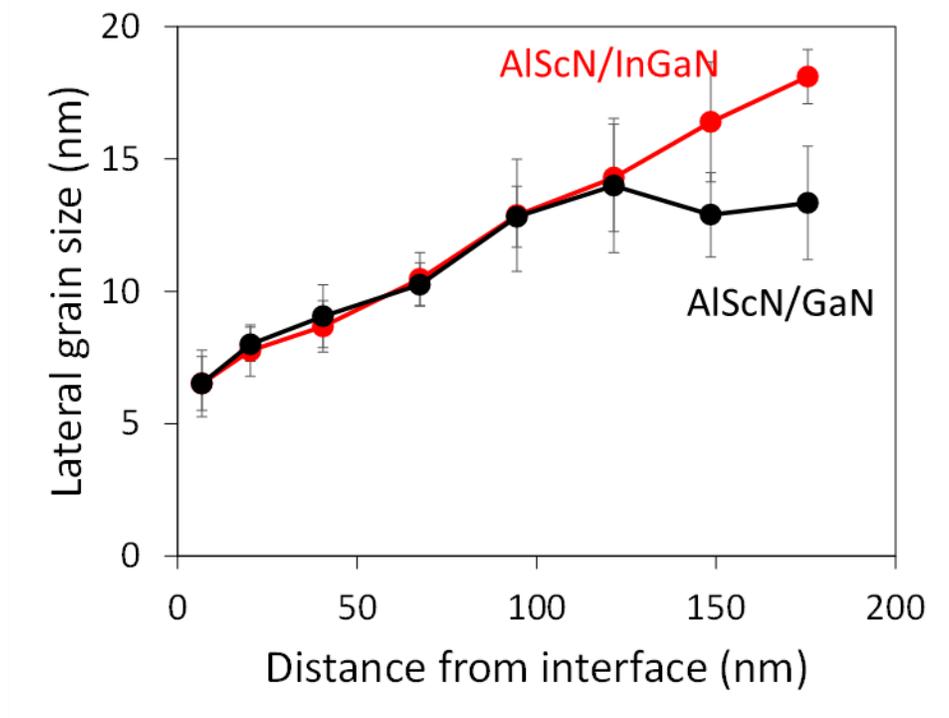

**Figure S6**. Lateral grain size obtained from differential phase contrast images of $Al_{0.7}Sc_{0.3}N$/GaN and $Al_{0.7}Sc_{0.3}N$/$In_{0.18}Ga_{0.82}N$. The film grown on $In_{0.18}Ga_{0.82}N$ shows linear growth of the crystallites with thickness. The film grown on GaN shows reduced crystallite growth at higher film thicknesses. Because of this difference the film grown on $In_{0.18}Ga_{0.82}N$ develops larger crystallites far from the interface: at distances greater than 100 nm from the interface the mean crystallite size in the film grown on $In_{0.18}Ga_{0.82}N$ is 16.3 nm, compared with only 13.4 nm in the film grown on GaN, although both films show similar crystallite sizes within 20 nm of the interface (6.8 nm on GaN versus 6.7 nm on $In_{0.18}Ga_{0.82}N$).

# Reference

[1]  E. A. Clinton, E. Vadiee, C. A. M. Fabien, M. W. Moseley, B. P. Gunning, W. A. Doolittle, A. M. Fischer, Y. O. Wei, H. Xie, and F. A. Ponce, Solid. State. Electron. **136**, 3 (2017).

[2]  M. Moseley, D. Billingsley, W. Henderson, E. Trybus, and W. A. Doolittle, J. Appl. Phys. **106**, (2009).

[3]  M. B. Tellekamp, M. K. Miller, L. Zhou, and A. Tamboli, J. Mater. Chem. C **11**, 13917 (2023).

[4]  S. R. Lee, A. M. West, A. A. Allerman, K. E. Waldrip, D. M. Follstaedt, P. P. Provencio, D. D. Koleske, and C. R. Abernathy, Appl. Phys. Lett. **86**, 1 (2005).

[5]  I. Lazić, E. G. T. Bosch, and S. Lazar, Ultramicroscopy **160**, 265 (2016).